\documentclass[aps,prl,twocolumn,superscriptaddress,nofootinbib,longbibliography]{revtex4-2}

\usepackage[utf8]{inputenc}
\usepackage{amsmath,amssymb}
\usepackage{graphicx}
\usepackage{hyperref}
\usepackage[dvipsnames]{xcolor}
\usepackage{bm}
\usepackage{longtable}
\usepackage{booktabs}
\usepackage[normalem]{ulem}
\usepackage{natbib}
\usepackage{diagbox}
\usepackage{multirow}
\usepackage{mathtools}
\usepackage[normalem]{ulem}

\providecommand{\aap}{A\&A}

\begin{document}

\title{
Exploring Cosmic Rays from Gravity-Limited Acceleration at \\
Supermassive Black Holes}

\author{Elena Manao}
\affiliation{Technical University of Munich, School of Natural Sciences, Physics Department, James-Frank-Str. 1, 85748 Garching bei München, Germany}

\author{Elisa Resconi}
\affiliation{Technical University of Munich, School of Natural Sciences, Physics Department, James-Frank-Str. 1,
85748 Garching bei München, Germany}

\author{Frank M. Rieger}
\affiliation{Max Planck Institute for Plasma Physics (IPP), Boltzmannstraße 2, 85748 Garching bei M{\"unchen}, Germany}
\affiliation{Institute for Theoretical Physics, Heidelberg University, Philosophenweg 12, 69120 Heidelberg, Germany}

\author{Paolo Padovani}
\affiliation{Technical University of Munich, School of Natural Sciences, Physics Department, James-Frank-Str. 1, 85748 Garching bei München, Germany}
\affiliation{European Southern Observatory, Karl-Schwarzschild-Straße 2, 85748 Garching bei M{\"u}nchen, Germany}

\author{Gia Dvali}
\affiliation{Arnold Sommerfeld Center, Ludwig-Maximilians-Universit{\"a}t, Theresienstr. 37, 80333 M{\"u}nchen, Germany}
\affiliation{Max-Planck-Institute for Physics, Boltzmannstr. 8, 85748 Garching bei M{\"unchen}, Germany}

\begin{abstract}
We show that horizon-scale magnetic fields around supermassive black holes (SMBHs), only limited by gravity, yield a universal, model-independent ceiling on electromagnetic energy density.
For rotating SMBHs, this field enables particle acceleration up to $\sim\!10^{27}$\,eV set solely by fundamental constants. 
We identify a parameter space in accretion rate and magnetic field strength in which quiescent nuclei simultaneously satisfy the transparency conditions for Ultra-High-Energy Cosmic Rays (UHECR) escape and the field requirements for reaching Ultra-High Energies (UHE), naturally complementing photon-rich Active Galactic Nuclei (AGN), which favor neutrino production but suppress UHECR escape.
We observe that, when including curvature energy losses, the energy of the primaries scales weakly as $E_{\rm curv}\!\propto\!M^{1/4}$ and allows to achieve UHE across different compositions and SMBHs masses ($10^6$--$10^{10}\,M_\odot$), without requiring extreme spins or super-Eddington accretion.
A cross-correlation of the Pierre Auger Observatory (PAO) public data with 
galaxies hosting reliably measured SMBH masses, separating AGN 
from quiescent hosts, reveals marginal local excesses 
($\lesssim\!2.8\sigma$) consistent with isotropy after 
correcting for the look-elsewhere effect.
The framework provides a direct link between black-hole physics and 
the maximum energies of cosmic particles, and defines concrete 
observational tests for next-generation UHECR datasets.
\end{abstract}

\maketitle
\textit{Introduction.---}
The IceCube Neutrino Observatory has reported evidence of neutrino 
emission from the blazar \mbox{TXS\,0506+056} and the X-ray-bright 
Seyfert galaxy NGC\,1068 
\cite{IceCube:2018ScienceFlare, IceCube:2018ScienceAlert, 
IceCube:2022Science}.
Taken together with hints in other blazars (e.g., PKS\,1424+240 \citep{Icecube:2025X-raybright}) 
and X-ray-bright AGN (e.g., NGC\,4151, CGCG\,420-015 
\citep{Icecube:2024Seyfert, Icecube:2025X-raybright}), these 
observations are consistent with astrophysical neutrinos originating 
in AGN, with the signal appearing agnostic to AGN 
type.
While in blazars, neutrinos are believed to originate in the jet 
\citep{Halzen:1997hw, Rodrigues2026}, for NGC\,1068 and other X-ray-bright AGN 
several studies favor efficient $p\gamma$ interactions of cosmic 
rays (CRs) accelerated within $\sim\!10$--$30\,R_G$ (gravitational radius) and 
interacting with dense coronal photon fields 
\cite{Inoue:2019yfs, Murase:2019vdl, Padovani:2024NatAstro}.
Such compact, photon-rich regions naturally explain the observed 
$\sim\!10$--$100$\,TeV neutrino fluxes, while 
rendering the escape of the parent CRs 
inefficient.
This suppression is further reinforced for high-mass nuclei, which are preferentially photodisintegrated in compact photon fields.
In summary, AGN that are efficient neutrino emitters are naturally disfavored as (U)HECR\footnote{NGC~1068 neutrinos require CRs of PeV energies.} sources.
These considerations prompt a central question:
\textit{if high-energy neutrinos trace dense, opaque AGN 
environments, where are the observed (U)HECRs accelerated?}
Because the conditions that enable strong electromagnetic acceleration in accretion-powered environments typically 
also imply strong radiative and plasma feedback suppressing primary escape, we move beyond 
accretion-based 
scaling arguments and 
examine a complementary scenario
 in which the near-horizon magnetic field is bounded by 
gravitational back-reaction alone: the \textit{gravity-limited} regime. In the conventional picture, the 
horizon-threading magnetic flux is regulated by the accretion state and poloidal field supply (e.g., 
Blandford--Znajek extraction), and this flux correlates with accretion luminosity across radio-loud AGN 
\cite{BlandfordZnajek1977, Tchekhovskoy2011, Zamaninasab2014}.
Here we probe the universal ceiling on the maximum magnetic field $B_{\rm max}$ to delineate the physically allowed parameter space for electromagnetic acceleration, analogous in spirit to Hillas-type confinement and radiative-loss bounds \cite{Hillas_1984, Aharonian2002}, and ask which aspects of the astrophysical environment survive at this extreme. Currently known systems operate at $B\!\ll\!B_{\rm max}$, with the realized field set by flux supply 
\cite{Aharonian2002}. In theory, the gravity-limited case defines the 
maximum possible acceleration voltage, which motivates a second question: \textit{does an astrophysical regime exist in which horizon-scale acceleration operates while the environment remains sufficiently transparent for (U)HECR escape?}
We address this by first deriving the gravity-limited magnetic field near black-hole 
horizons, and establishing the maximal attainable energy for particles in these environments.
Then, we identify the transparency conditions (accretion rate, photon density, and pair loading) under which quiescent nuclei simultaneously 
sustain strong fields and allow (U)HECR escape. In particular, we focus on the highest detected energies by testing whether PAO UHECR arrival directions correlate with galaxies hosting SMBHs with reliably measured masses \cite{Kormendy_2013, Saglia_2016}, separating AGN from quiescent hosts characterized by low Eddington ratios\footnote{Defined as the ratio between the bolometric luminosity and the Eddington luminosity ($L_{\rm Edd} = 1.26\cdot10^{46}\, (M/10^8\,M_\odot)$\,erg\,s$^{-1}$).}, as motivated by the aforementioned parameter space.

\textit{Maximum magnetic field from gravity-limited constraints.---}
We impose that the horizon scale is comparable in size to the Schwarzschild radius ($R_S$), meaning $R\!\sim\!R_S\!=\!2GM/c^2$, with $G$ the 
gravitational constant, $c$ the speed of light, and $M$ the SMBH 
mass, and require that the magnetic energy density remains 
subdominant to gravity,
\begin{equation}
\frac{B^2}{8\pi} \lesssim \rho_{\rm grav} \sim \frac{GM^2}{R^4},
\end{equation}
where $\rho_{\rm grav}$ is the characteristic 
gravitational energy density on scale $R$, obtained by distributing 
mass $M$ over volume $R^3$ with curvature scaling as $GM/R$.
This constraint yields a universal upper ceiling on the magnetic field
\cite{Dvali2022, Dvali:2025mqr}
\begin{equation}\label{eq:Bmax}
B_{\rm max} \sim \frac{c^4}{G^{3/2}M},
\end{equation}
up to $O(1)$ factors, representing a 
fundamental, model-independent bound set by gravity alone, 
applicable to any horizon-scale magnetic configuration. 
In natural units, with Planck mass 
$M_P\!=\!1/\sqrt{2G}$ and $R\!\propto\!GM$, 
$B_{\rm max}$ can be rewritten as
\begin{equation}\label{eq:Bmax_quantum}
B_{\rm max} \sim \frac{M_P}{R},
\end{equation}
which admits a complementary quantum interpretation: the occupation 
number of a coherent classical magnetic field in a region of size 
$R$ scales as $N_\gamma\!\sim\!B^2R^4$ and cannot exceed the 
Bekenstein--Hawking entropy $S_{\rm BH}\!\sim\!(M_PR)^2$ 
\cite{Dvali:2011aa, Dvali2022}.
Equating the two recovers Eq.~(\ref{eq:Bmax}), 
confirming the bounds' universality from both classical GR$+$EM 
and black-hole entropy arguments.
For a $10^9\!M_\odot$ SMBH, 
$B_{\rm max}\!\sim\!10^{10}$\,G, well below the Schwinger field, the quantum-electrodynamic critical field ($4.4\!\cdot\!10^{13}$\,G) at which vacuum pair production becomes efficient, and far above observationally inferred horizon-scale fields: $\sim\!40$--$90$\,G in flaring Sgr~A* \cite{Gravity_sgrA_2021, flaring_sgrA_2013} and $\sim\!10^4$--$10^5$\,G near AGN jet bases \cite{BlandfordZnajek1977, Yuan2014, Katsoulakos2018}.
Thus, $B_{\rm max}$ is not a prediction for any specific source 
but a robust, model-independent ceiling on the near-horizon 
electromagnetic energy density: the starting point for the 
parameter space analysis that follows.

\textit{Luminosity constraint.---}
While the gravity-limited ceiling sets the maximal acceleration voltage, 
UHECR escape also requires a sufficiently transparent environment, which constrains the luminosity and photon density of the source.
In radiatively efficient AGN, dense soft-photon fields quench magnetospheric gaps via pair cascades and suppress UHECR escape via $p\gamma$ interactions \cite{Levinson2011, Broderick2015, Hirotani2016}.
We therefore focus on low-luminosity nuclei with radiatively inefficient 
accretion flows (RIAF) \cite{Narayan1994, Yuan2014}, for which the 
luminosity scales as
\begin{equation}
L \propto \dot{m}^2 L_{\rm Edd},
\end{equation}
We define \textit{quiescent} nuclei as systems with accretion rate
$\dot{m} \lesssim 10^{-4}$,
for which photon-induced losses and gap screening are subdominant 
(see Supplemental Material (SM)~\cite{SM}).

\textit{Magnetic field constraints.---}
The gravity-limited bound derived above represents a universal upper ceiling on horizon-scale magnetic fields. In known systems, the horizon-scale field is set by the balance between magnetic pressure and the ram pressure of the infalling plasma, and therefore scales with the 
accretion rate. Equipartition arguments for RIAF imply inner fields of order \cite{Narayan1995, Yuan2014}
\begin{equation}\label{eq:B_equip}
B_{\rm eq} \sim 10^{5}\,\dot{m}^{1/2}\,M_9^{-1/2}\ {\rm G},
\end{equation}
with $M_9 = M_{\rm SMBH}/10^9\,M_\odot$ (see SM for more details \cite{SM}). This is significantly below the 
gravity-limited ceiling $B_{\rm max}$ (Eq.~\ref{eq:Bmax}).
Furthermore, once the accumulated 
magnetic flux approaches the magnetically arrested disk (MAD) threshold 
\cite{Narayan2003,Tchekhovskoy2011}, magnetic pressure halts further 
flux accumulation, preventing sustained fields significantly above 
$B_{\rm eq}$. Thus, while $B_{\rm max}$ sets a universal ceiling, 
quiescent nuclei naturally realize sub-maximal fields $B_{\rm eq} \ll 
B_{\rm max}$.

\textit{Gap width and pair loading.---}
A horizon-scale potential drop requires that the accelerating region 
is not screened by pair cascades \cite{Levinson2011, Hirotani2016}.
A necessary condition is that curvature or inverse-Compton photons 
emitted by accelerated particles do not undergo efficient 
$\gamma\gamma\!\to\!e^+e^-$ conversion on the ambient soft radiation 
field over a distance shorter than a gap size $R_{S}$.
The pair-production mean free path is
$\ell_{\gamma\gamma} \simeq \frac{1}{n_\gamma\,\sigma_{\gamma\gamma}}$,
where the $\gamma\gamma$ cross-section is $\sigma_{\gamma\gamma}\!\sim\!10^{-25}\,\mathrm{cm}^2$\cite{Zdziarski1988}.
Using the soft-photon density expected in quiescent nuclei 
(SM \cite{SM}), one finds
\begin{equation}
\ell_{\gamma\gamma} \sim 3\cdot10^{14}\,
M_9\,\hat{R}^{\,2}\,
\left(\frac{10^{-4}}{\dot{m}}\right)^{2}
\left(\frac{\epsilon}{1\,\mathrm{eV}}\right)\,\mathrm{cm},
\end{equation}
where $\hat{R}\!\coloneqq\!R/R_S$ and $\epsilon$ is 
the target photon energy. 
Given that $\ell_{\gamma\gamma}\!\simeq\! R_S$ for $\dot{m}\!\sim\!10^{-4}$, quiescent nuclei admit gap width $\gtrsim\!R_S$.
In contrast, in photon-rich environments $\ell_{\gamma\gamma}\!\ll\!R_S$ 
and gaps are quenched \cite{Levinson2011}.
The relevant target-photon energies depend on the Lorentz factor of the emitting particles and are discussed in the SM \cite{SM}.

\textit{Photodisintegration.---}
For heavy nuclei, escape additionally requires that photodisintegration (PD) in the near-horizon radiation field is suppressed.
Analogously to the pair-production case just discussed, the PD interaction length,
\begin{equation}
\ell_{\rm PD}\simeq\frac{1}{n_\gamma\,\sigma_{\rm PD}},
\end{equation}
must be $\ell_{\rm PD}\!\gtrsim\!R_S$ for nuclei to escape.
Near the giant-dipole resonance (GDR), $\sigma_{\rm PD}\sim\!10^{-25}\,\mathrm{cm}^2$ \cite{Puget1976,Khan2005}, 
comparable to $\sigma_{\gamma\gamma}$ at its peak, so $\ell_{\rm PD}\sim\ell_{\gamma\gamma}$ for the same target photon density.
The same condition $\dot{m}\!\lesssim\!10^{-4}$ that ensures gap survival also suppresses photodisintegration in quiescent nuclei, providing a unified transparency window for both gap maintenance and heavy-nuclei escape.
In contrast, photon-rich environments efficiently photodisintegrate UHE nuclei well within $R_S$, consistent with the exclusion of radiatively efficient AGN as UHECR sources.

\textit{Poynting-flux luminosity.---}
For a rotating force-free Kerr BH magnetosphere, the Blandford--Znajek 
(BZ) power is \cite{BlandfordZnajek1977}
\begin{equation}
L_{\rm BZ} \simeq \Omega_F(\Omega_H - \Omega_F)\,
\frac{B_\perp^2 R_H^{4}}{c}
\simeq \frac{a^{2}}{16}\,c\,R_H^{2}B_\perp^{2},
\label{eq:LBZ_main}
\end{equation}
where $R_H = R_S(1+\sqrt{1-a^2})/2$ is the horizon radius, $a \leq 1$ the 
dimensionless spin, $B_\perp$ the poloidal field threading the horizon, 
$\Omega_H = a c/(2r_H)$ the horizon angular velocity, with efficient extraction 
corresponding to field line rotation $\Omega_F \simeq \Omega_H/2$.
Substituting the gravity-limited ceiling $B_\perp \lesssim B_{\rm max}$ 
formally yields
$ L_{\rm BZ} \sim (a_s^2/16)\,(c^5/G) 
\sim 10^{58}\,a_s^2\ \mathrm{erg\,s^{-1}}$,
i.e., of order the Planck luminosity $L_P = c^5/G \simeq 3.6\cdot10^{59}$\,
erg\,s$^{-1}$ \cite{Dvali2022}, a direct consequence of $B_{\rm max}$ being 
set by the same gravitational scale. While we do not expect a steady field 
as strong as $B_{\rm max}$ to be realized, we note that this estimate  
assumes a quasi-steady, force-free, current-carrying magnetosphere 
\cite{BlandfordZnajek1977,Tchekhovskoy2011}. In the absence of sufficient 
plasma to sustain the required currents, as might occur in quiescent BHs, 
the magnetosphere may deviate from the force-free regime, and the resulting 
Poynting luminosity can fall well below the force-free BZ maximum.
In principle, larger voltages in such charge-depleted gaps could therefore 
coexist with weak jet and radiative output. 

\textit{Geometric Energy Limit from Horizon-Scale Fields.---} Magnetic fields bend trajectories, but acceleration requires an electric component along the field.
Near rotating black holes, frame dragging, and magnetospheric gaps can provide such $E_\parallel$ (as in pulsars~\cite{Ruderman1975, Hirotani_radiation_rotating_SMBH}, M87~\cite{Levinson2000,Levinson2011, Aharonian2006}, and the Galactic Centre~\cite{HESS2016Nature, Rieger_acc_SGRA}).
For a rotating SMBH threaded by a magnetic field, 
we parametrize the maximum energy gain geometrically by considering a potential drop over the horizon scale~\cite{Hillas_1984}, analogous in spirit to the Hillas confinement criterion, but here arising from unscreened near-horizon electric fields generated by frame dragging rather than from confinement arguments, yielding
\begin{equation}\label{eq:Egeom}
E_{\rm geom} = a\,Ze\,B\,R_S,
\end{equation}
with charge $Ze$.
Substituting $B_{\rm max}$ (Eq.~(\ref{eq:Bmax})) gives
\begin{equation}\label{eq:massfree}
E_{\rm geom} = 2\,a\,Ze\,\frac{c^2}{\sqrt{G}}
\;\simeq\; 10^{27}\,Za\ \mathrm{eV},
\end{equation}
a mass-independent ceiling set solely by 
fundamental constants and the black-hole spin, independent of 
the SMBH mass, accretion state, or environment.
It is intriguing to see that quantum gravity considerations arrive at the same ceiling: viewing the horizon-scale field as a flux tube of coherence length $R\,\sim R_S$, its transverse energy per unit length scales as $\mu\!\sim\!(BR)^2$ and saturates at $\mu\!\sim\!M_P^2$ when $B\!\to\!B_{\rm max}$ (Eq.~(\ref{eq:Bmax_quantum})), the maximum compatible with asymptotically flat geometry.
 
The largest energy gain for a particle of charge $q\!=\!Ze$ traversing the flux tube is then $
E_{\rm max} \sim q\sqrt{\mu} \sim qM_P$,
matching Eq.~(\ref{eq:massfree}) in natural units for $a\!\sim\!1$. For a sub-maximal $B$, the attainable energy scales linearly,
\begin{equation}
E_{\rm max} \sim qM_P\left(\frac{B}{B_{\rm max}}\right),
\end{equation}
recovering the expected proportionality to the available magnetic flux. 
The energies achievable in realistic astrophysical environments are determined by the actual field strength $B_{\rm eq}\!\ll\!B_{\rm max}$ and by radiative losses, both addressed in the following sections.

\begin{figure*}[ht]
    \centering
    \includegraphics[width=\linewidth]{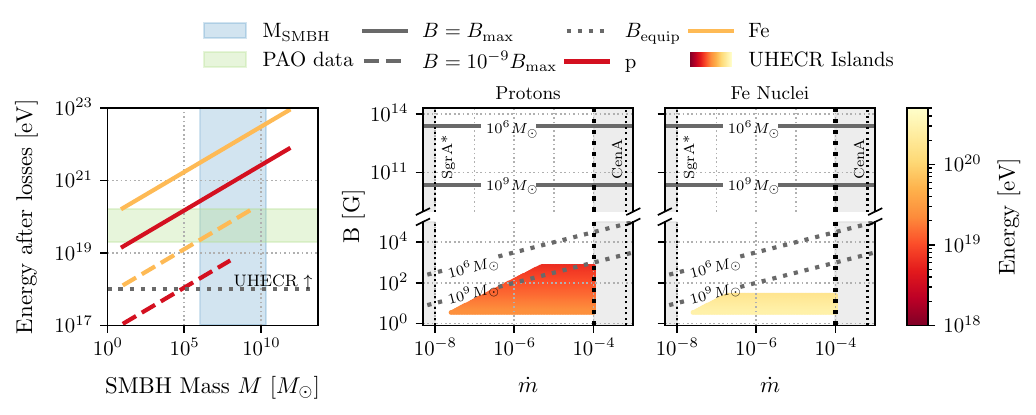}
    \caption{
    Left: Curvature-limited energies for protons (p, red) and iron nuclei (Fe, orange) as functions of $M_{\rm SMBH}$ for different magnetic field strengths and $a\!\sim\!1$. We highlight the range of SMBH masses considered in this work (blue, see next section) and that of UHECR energies used in the analysis (green). The lines extend only as long as $E_{\rm geom}$ (Eq.~\ref{eq:Egeom}) is larger than $E_{\rm curv}$ (Eq.~\ref{eq:Ecurv_gen}), which results in lower maximum energies being reached at lower $B$ (and therefore higher masses). 
    Center and right: Allowed parameter space, i.e., the \emph{UHECR island}, for horizon-scale UHECR acceleration in quiescent galaxies covering our sample's mass range for $B$ values compatible with the ones attainable with equipartition arguments ($B_{\rm equip}$, Eq.\,\ref{eq:B_equip}), calculated for the representative case of $B=10^{-9}B_{\rm max}$. The color scale represents the range of particles' energy. Note that protons reach energies up to $3\!\cdot\!10^{19}{\rm eV}$ (darker colors), while iron nuclei reach $O(10^{20} {\rm eV})$ energies (light colors).
    The solid lines show $B_{\rm max}$ (Eq.~\ref{eq:Bmax}) while the dotted lines show $B_{\rm eq}$ (Eq.~\ref{eq:B_equip}) for two representative values of $M_{\rm SMBH}$. 
    Vertical dotted lines indicate, from left to right, the estimated accretion rates of  Sgr~A* ($\dot{m}\!\sim\!10^{-8}$), treated as a reference for sources with an extremely diluted disk \cite{Gravity_sgrA_2021}, the accretion values for which the environment is considered transparent ($\dot{m}\!\lesssim\!10^{-4}$), and Cen~A ($\dot{m}\!\sim\!10^{-3}$), the most promising AGN candidate identified up to now.}
    \label{fig:B_Emax_vs_M}
\end{figure*}

\textit{Energy losses and critical mass range.---}
In strong magnetic fields, accelerated charges rapidly lose transverse momentum, 
effectively sliding along the field, with curvature radiation setting the
relevant loss ceiling \cite[e.g.,][]{Rieger2011, Blumenthal1970}.
Balancing the acceleration rate, estimated as the energy gain over one 
horizon-crossing time, with the curvature loss rate \cite{Jackson1975},
\begin{equation}
\left(\frac{dE}{dt}\right)_{\rm gain} = Ze\,a\,B\,c
\;=\;
\left(\frac{dE}{dt}\right)_{\rm loss} = 
\frac{2}{3}\frac{(Ze)^2 c\,\gamma^4}{R_c^2},
\end{equation}
with Lorentz factor $\gamma$ and curvature radius $R_c$, yields the 
curvature-limited energy
\begin{equation}\label{eq:Ecurv_gen}
E_{\rm curv} = \left(\frac{3\,a\,m^4c^8\,B\,R_c^2}{2\,Ze}
\right)^{1/4}.
\end{equation}
Adopting the horizon scale as characteristic curvature radius, $R_c\!\sim\!R_S$, and substituting $B_{\rm max}\!\propto\!M^{-1}$ and $R_S\!\propto\!M$ 
from Eqs.~(\ref{eq:Bmax})--(\ref{eq:Egeom}), gives
\begin{equation}\label{eq:Ecurv_scaling}
E_{\rm curv} \propto 
\left(B_{\rm max}\,R_S^2\right)^{1/4} \propto
M^{1/4},
\end{equation}
so that more massive black holes reach higher 
loss-limited energies.
For a $10^9\,M_\odot$ SMBH with $a\!\sim\!1$, Eq.~(\ref{eq:Ecurv_gen}) yields benchmark values $E_{\rm curv}\!\sim\!10^{21}$\,eV for protons and $\sim\!10^{22}$\,eV for iron as shown in Fig.~\ref{fig:B_Emax_vs_M}.
The associated curvature photons, with characteristic energy $\epsilon_{\rm curv}\!\sim\!\frac{3}{2}\hbar c\,\gamma^3/R_c$, extend to multi-PeV energies and are subject to $\gamma\gamma$ absorption on local radiation fields and pair production on the extragalactic background light and cosmic microwave background during propagation. 

Since $E_{\rm geom}$ has to be $> E_{\rm curv}$ and given their different scalings ($\propto B$ and $\propto B^{1/4}$ respectively: cf. Eq.~\ref{eq:Egeom} with Eq.~\ref{eq:Ecurv_gen}) at very low $B$ values the particle energies are slightly reduced compared to what predicted by Eq.~\ref{eq:Ecurv_gen}. 
However, both light and heavy nuclei can still reach UHECR energies across the relevant SMBH mass range (Fig.~\ref{fig:B_Emax_vs_M} left).
In Fig.~\ref{fig:B_Emax_vs_M} right, we show that magnetic field equipartition values (SM, eq. (7)), which are many orders of magnitude smaller than the gravity-limited value (eq. \ref{eq:Bmax}), still allow for the production of UHECRs even for quiescent galaxies hosting SMBHs. 
The allowed parameter space identified in Fig.~\ref{fig:B_Emax_vs_M} motivates the source selection considered next.
Massive, radiatively inefficient (observationally 
quiescent) galactic nuclei thus emerge as natural UHECR accelerator candidates within this framework.

\begin{figure}
    \centering
    \includegraphics[width=.9\linewidth]{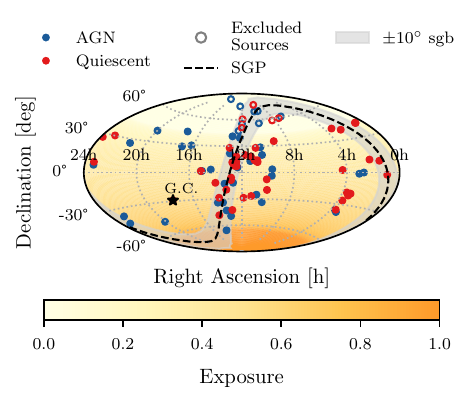}
    \caption{Sources: blue = AGN, red = quiescent galaxies, open symbols = outside the PAO field of view. The dashed line marks the supergalactic plane (SGP); the shaded band shows $|\mathrm{sgb}|<10^\circ$. The background shows the PAO normalized exposure. The galactic center (G.C.) is marked with a star.}
    \label{fig:sources}
\end{figure}

\textit{Galaxy sample selection.---}
To probe the potential of SMBHs as UHECR accelerators \cite{Torres2002}, we assemble 
a sample of galaxies with directly measured SMBH masses from 
\cite{Kormendy_2013} and \cite{Saglia_2016}, yielding 103 systems 
spanning $10^{6}$--$10^{10}\,M_\odot$ at distances 
$0.77$--$242$\,Mpc.
No distance cut was imposed; the upper bound 
reflects the angular resolution limit for resolving the SMBH sphere 
of influence with current instrumentation \cite{Padovani_ELT}.
We classify sources as \textit{active} (AGN; $N\!=\!49$) when recent 
literature reports accretion signatures (jets, hard X-rays, or 
$\gamma$-rays), and as \textit{quiescent} nuclei otherwise 
($N\!=\!54$).

The sky distribution is anisotropic, with 45 objects concentrated along the supergalactic plane ($|\mathrm{sgb}|\!<\!10^\circ$), consistent with the local large-scale structure (Fig.~\ref{fig:sources}).
This sample is well matched to UHECR studies: the source distances 
lie within the expected propagation horizons for both light and heavy 
primaries \cite{Attenuation}, and the SMBH masses are sufficient, 
within our gravity-limited framework, to sustain acceleration to the 
energy range probed here (Fig.~\ref{fig:B_Emax_vs_M}).
We then restrict the sample to the PAO field of view ($-90^\circ\!\leq\!\delta\!\leq\!44.8^\circ$), yielding 90 sources: 41 AGN and 49 quiescent hosts.
The Galactic Center (GC),
is excluded from the analysis due to its low $\dot{m}$ \cite{Rieger_acc_SGRA}. 
However, we also perform the same test including the GC as a cross-check. The results are presented in the SM \cite{SM}.

\textit{Correlation with Pierre Auger Data.---}%
\footnote{Analysis code available at 
\href{https://github.com/elenamanao/smbh/tree/main}{this GitHub 
repository}.}
\begin{figure}[t]
    \centering
    \includegraphics[width=.9\linewidth]{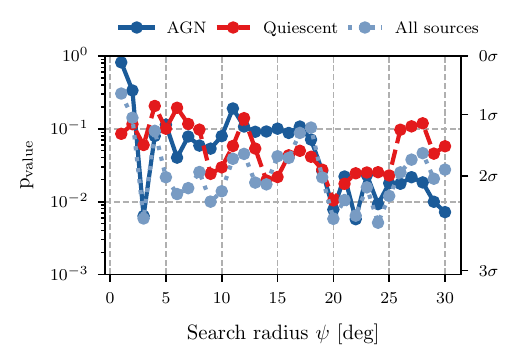}
    \caption{Local correlation significance vs.\ 
    search radius $\psi$ for AGN (solid blue), quiescent nuclei 
    (dashed red), and the full sample (dotted light blue), expressed 
    as both p-value and equivalent Gaussian significance $\sigma$.}
    \label{fig:results}
\end{figure}
We test for spatial correlations between our 90-source sample (41 AGN, 49 quiescent nuclei) and the PAO public dataset of \citet{PAO_Anisotropy2022_ApJ} \citep{PAO_data_release}, comprising $N\!=\!2635$ events with $E\!\geq\!32$\,EeV recorded from 2004 to 2020 
($2040$ vertical events, 
$0^\circ\!\leq\!\theta\!\leq\!60^\circ$; 595 inclined events, 
$60^\circ\!\leq\!\theta\!\leq\!80^\circ$).
For each source class (AGN, quiescent nuclei, combined), we count events within angular radius $\psi$ and compare the associated-event fraction to isotropic simulations weighted by the PAO declination-dependent exposure and the measured large-scale anisotropy \citep{pao_anisotropy_2024} (details in SM~\cite{SM}).We scan $\psi\!=\!1^\circ$--$30^\circ$ in $1^\circ$ steps, following the strategy of previous PAO analyses on the public dataset \cite{PAO_first_arrival_dir_analysis, 2018_anisotropy_apjl, PAO_Anisotropy2022_ApJ}, and covering the range of plausible magnetic deflections for UHECRs \cite{deflections:2021PRD}.

\textit{Results.---} Local p-values, prior to correction for multiple 
tested hypoteses, are shown in Fig.~\ref{fig:results}.
The most significant local excess occurs for the combined sample at $\psi\!=\!24^\circ$ (p\,$=5.1\!\cdot\!10^{-3}$, $2.79\sigma$); for AGN at $\psi\!=\!22^\circ$ (p\,$=5.8\!\cdot\!10^{-3}$, $2.76\sigma$); and for quiescent nuclei at $\psi\!=\!20^\circ$ (p\,$=1.0\!\cdot\!10^{-2}$, 
$2.56\sigma$). These local significances do not account for scanning three source classes over 30 angular radii.
After correcting for this look-elsewhere effect via the distribution of minimum p-values across all tested hypotheses in isotropic pseudo-experiments (SM~\cite{SM}), the most significant excess reduces to $1.25\sigma$, consistent with isotropy.
We note a secondary excess at $\psi\!=\!3^\circ$, 
driven by a small number of events in the close vicinity of 
individual sources; this is discussed further in the SM~\cite{SM}. The absence of a significant excess does not exclude the considered acceleration framework: given the current dataset size, the expected signal-to-noise for the source densities and deflection angles considered here remains marginal, and the analysis defines a concrete test for future PAO and next-generation UHECR datasets.

\begin{figure}[t]
    \centering
    \includegraphics[width=0.9\linewidth]{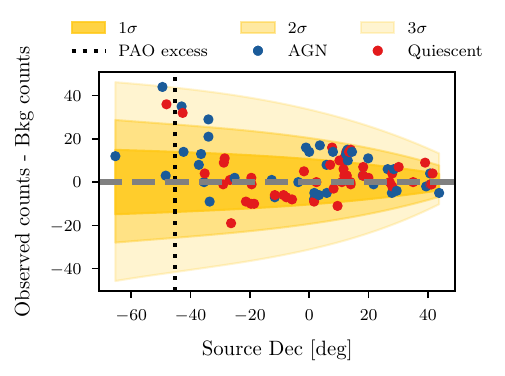}
    \caption{Per-source excess relative to the median 
    background for quiescent nuclei (red) and AGN (blue). 
    Shaded bands: $1\sigma$, $2\sigma$, $3\sigma$ background 
    contours. Three AGN (NGC\,4945, Cen\,A, NGC\,5419) and two 
    quiescent nuclei (NGC\,5516, NGC\,4751) 
    exceed $2\sigma$. The dashed line marks the declination of 
    the most significant PAO excess 
    ($\delta\!=\!-45.4^\circ$)~\cite{PAO_SGP_2025}.}
    \label{fig:bkg_deviation}
\end{figure}

\textit{Per-source view.---}
Although globally non-significant, five objects show 
$>\!2\sigma$ local excesses (Fig.~\ref{fig:bkg_deviation}): 
among AGN, NGC\,4945 ($\delta\!=\!-49.5^\circ$), Cen\,A 
($-43.0^\circ$), and NGC\,5419 ($-34.0^\circ$); among quiescent 
nuclei, NGC\,5516 ($-48.1^\circ$) and NGC\,4751 ($-42.6^\circ$).
All five lie in the Centaurus region, within $2^\circ$--$13^\circ$ 
of the best-fit coordinates of the most significant PAO excess at (r.a., $\delta$) $=$ (13.4\,h, $-45.4^\circ$) 
with radius $27^\circ$ \cite{PAO_SGP_2025}.
Of these, only Cen\,A and NGC\,4945 had been 
previously identified as UHECR source candidates 
\cite{PAO_Anisotropy2022_ApJ}; the remaining three objects, 
including two quiescent nuclei, overlap with the PAO excess region and are consistent with the gravity-limited acceleration framework presented here, which does not restrict UHECR production to radiatively active sources.

\textit{Conclusions.---}
We have explored the implications of large magnetic fields up to the gravity-limited regime for UHECR acceleration, identifying maximum achievable particle energies exceeding the UHECR threshold.
When curvature losses are included, the attainable energy scales weakly as $E_{\rm curv}\!\propto\!M^{1/4}$ and remains sufficient for both protons and heavy nuclei across $10^6$--$10^{10}\,M_\odot$, without invoking radiatively bright accretion or extreme spins.
Crucially, the mechanism operates locally at the horizon and does not require strong relativistic jets, which depend on the global accretion state, while the maximum attainable energy is set solely by horizon-scale fields. 
Guided by this framework, we have identified a parameter space in $(\dot{m}, B)$ (Fig.~\ref{fig:B_Emax_vs_M}) in which quiescent nuclei simultaneously satisfy the transparency conditions for particle escape and the field requirements for UHECRs. Our cross-correlation of PAO events with galaxies hosting reliably measured SMBH masses reveals marginal local excesses at the level of $\sim\!2.5$--$2.8\sigma$ before the look-elsewhere correction, which are not significant after accounting for multiple tests, reducing to $1.25\sigma$ and remaining consistent with isotropy.
Despite inconclusive statistical results, the decisive ingredients to understand UHECR appear to be the presence of a massive SMBH and local conditions enabling unscreened acceleration and particle escape, conditions likely met in quiescent nuclei.
Current statistics and magnetic deflections blur class contrasts, and a mixed-origin scenario remains consistent with the data.
We note that the suppression in the UHECR spectrum observed at $\sim\!5\cdot10^{19}$\,eV is consistent with GZK energy losses during propagation \cite{Greisen1966, Zatsepin1966} 
and does not necessarily imply a source acceleration ceiling; the maximum energies derived here comfortably exceed this threshold.
Looking ahead, on the observational side, larger datasets combined with rigidity-aware likelihoods, energy--angle weighting, and source-distance/mass priors will sharpen sensitivity.
The Centaurus region sources identified here (Cen\,A, NGC\,4945, NGC\,5516, NGC\,4751, NGC\,5419) are concrete targets for follow-up with next-generation UHECR observatories. On the 
theoretical side, quantifying gap formation, field topology and cosmic-ray injection in quiescent nuclei, and testing the role of atomic nuclei coalescence near the horizon are key 
to resolving the heavy-nucleus UHECR component and the complementarity of UHECR accelerators and neutrino emitters.

\textit{Acknowledgments.—}
We thank the anonymous referees for their insightful comments on the first version of this manuscript. We thank Chiara Bellenghi, Ruohan Li, Zaza Osmanov, Taro Shimizu, and Sebastiano von Fellenberg for valuable discussions. E.M. is funded by the Deutsche Forschungsgemeinschaft (DFG, German Research Foundation) via SFB~1258 “Neutrinos, Dark Matter, and the Messengers of the Universe” (Project-ID 283604770). The work of G.D. was supported in part by the Humboldt Foundation under Humboldt Professorship Award, by the European Research Council Gravities Horizon Grant AO number: 850 173-6, by the Deutsche Forschungsgemeinschaft (DFG, German Research Foundation) under Germany's Excellence Strategy - EXC-2111 - 390814868, and Germany's Excellence Strategy under Excellence Cluster Origins. We also acknowledge support from the ORIGINS Cluster of Excellence (Connector~6). AI tools were used to assist with figure aesthetics and text drafting; all outputs were critically reviewed and verified by the authors to ensure accuracy and integrity.

\textit{Disclaimer.—} Funded by the European Union. Views and opinions expressed are however those of the authors only and do not necessarily reflect those of the European Union or European Research Council. Neither the
European Union nor the granting authority can be held responsible for them. 

%

\null
\thispagestyle{empty}
\clearpage

\onecolumngrid
\setcounter{section}{0}
\setcounter{equation}{0}
\setcounter{figure}{0}
\setcounter{table}{0}
\renewcommand{\thesection}{S\Roman{section}}
\renewcommand{\theequation}{S\arabic{equation}}
\renewcommand{\thefigure}{S\arabic{figure}}
\renewcommand{\thetable}{S\arabic{table}}
\renewcommand{\theHequation}{S\arabic{equation}}
\renewcommand{\theHfigure}{S\arabic{figure}}
\renewcommand{\theHtable}{S\arabic{table}}

\begin{center}
{\Large\bfseries Supplementary Material}
\end{center}

\section{Physical Constraints} 
In the following, we provide basic considerations regarding the acceleration and escape of UHECRs 
from quiescent hosts.

\subsection{Luminosity and Photon Density Constraints}
In order for horizon-scale electromagnetic acceleration to operate efficiently, the surrounding radiation 
field must remain sufficiently dilute such that photon-induced interactions do not attenuate accelerated 
particles and pair creation does not screen the accelerating electric field. 
We define \emph{radiatively inefficient accretion} (RIA) as accretion at rates for which the flows becomes 
radiatively inefficient. In this case, cooling is inefficient and the bolometric luminosity ($L_{\rm bol}$) 
scales approximately as
\begin{equation}
L_{\rm bol} \sim \dot m^2 L_{\rm Edd},
\end{equation}
where $\dot m=\dot M/\dot M_{\rm Edd}\ll1$ is the dimensionless accretion rate and $L_{\rm Edd}  
\simeq 1.3\times10^{47} M_9 \ \mathrm{erg\,s^{-1}}$ is the Eddington luminosity, with $M_9 := 
M_{\rm BH}/10^9 M_\odot$.
Under such conditions, the reference \emph{photon number density} at characteristic photon energy 
$\epsilon_\gamma$, assuming isotropic emission at radius $R=\hat R R_S$ (where the Schwarzschild 
radius is $R_S = \frac{2GM_{\rm BH}}{c^2}\simeq 3\times10^{14} M_9 \ \mathrm{cm}$) becomes
\begin{equation}\label{eq:riaf_number_density}
n_\gamma = \frac{L_{\rm bol}}{4\pi r^2 c \epsilon_\gamma} 
\simeq 3\times10^{18}\frac{\dot m^2}{M_9 \hat R^2}\left(\frac{1\,\mathrm{eV}}{\epsilon_\gamma}\right)
\mathrm{cm^{-3}} 
\propto \dot m^2 M_9^{-1} \hat R^{-2}.
\end{equation}
highlighting its scaling with the BH mass, accretion rate and source size in units of $R_S$. The 
quadratic dependence on $\dot m$ implies a rapid decrease in interaction opacity as the accretion 
rate declines. For sufficiently small $\dot {m} $, the pair-production and photodisintegration 
timescales exceed the dynamical escape time $t_{\rm esc}\sim R_S/c$, allowing the formation of 
macroscopic gaps and the escape of accelerated particles \citep[e.g.,][]{Rieger2011}.
Adopting representative photon energies of order eV and evaluating interaction optical depths near 
$r\sim(1$--$10)\,R_S$, we find that photon-induced losses and gap screening become subdominant for
$\dot m \lesssim 10^{-4}$, which motivates the definition of \emph{quiescent or passive} sources 
used in the main text.

\subsection{Magnetic Field Constraints}
The current work considers particle acceleration near the event horizon where the acceleration
efficiency scales with magnetic field strength. Because stationary currents cannot exist on the
horizon itself, the field must be supported by current in the surrounding plasma. If accretion 
were to cease, these currents would disappear and the magentic field decay via reconnection and 
diffusion on timescales $t_r \sim (10\text{--}100)\, R_G/c$ as expected from general considerations 
\citep{Price1972,Thorne1986} and confirmed by recent numerical simulations \citep{Bransgrove2021}.
In realistic astrophysical environments, equipartition arguments (i.e., equating the magnetic-field
energy density to the ram pressure of the inflowing plasma) imply that the horizon-scale field 
strength in magnetized accretion flows is of order \citep[e.g.,][]{Yuan2014,Katsoulakos2018}
\begin{equation}\label{eq:B_eq}
B_{\rm eq} \simeq 10^5 \,\dot m^{1/2} M_9^{-1/2} {\rm G}\,.
\end{equation}
For low accretion rates ($\dot{m} \lesssim 10^{-4}$) this yields typical field strengths $B_{\rm eq} 
\lesssim 10^3\, M_9^{-1/2}\ {\rm G}$, which can be large enough to satisfy the requirements for UHECR 
acceleration. The equipartition estimate (Eq.~\ref{eq:B_eq}) is equivalent to the Lorentz force
density $f_B \sim B^2/(8\pi R_s)$ not exceeding the gravitational force density $f_g \sim (1/2) n 
m_p c^2 (R_s/R^2)$, i.e., $B \lesssim 0.1 n^{1/2}$ G, with $n$ the plasma density in cm$^{-3}$. 
While in principle even a modest global toroidal current can sustain a large poloidal magnetic 
field ($B_p \sim 4 \pi R_s j_t/c$ by Ampère's law), the associated magnetic pressure will suppress
accretion, making it difficult to maintain a steady, large-scale field in astrophysical systems. 
Thus, a steady, gravity-limited magnetic field of strength $B_{\rm max}$, as discussed in the main 
text, is unlikely to be realized. One could image that accretion might continue through intermittent 
reconnection events, turbulence, or by channeling the gas along field lines, potentially permitting 
transiently stronger fields, but such possibilities are not explored further here.

\section{Simulation Procedure}\label{app:simulation}
\begin{figure}[h]
    \centering
    \includegraphics[width=0.75\textwidth]{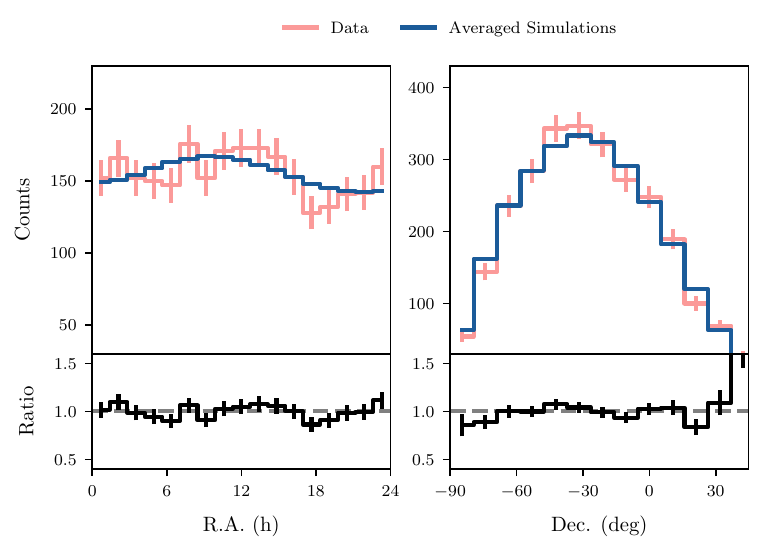}
    \caption{Comparison between data (pink) and 1000 averaged simulation (blue). On the left, we show the R.A. distribution, on the right, the declination. In the lower panels, we display the ratio between the data and the simulation, which averages around 1 over the entire coordinate range.}
    \label{fig:simulation}
\end{figure}

The significance of the results presented in this work is determined by comparing the outcome of the statistical tests with simulations. To produce reliable, robust results, the simulations must provide a good description of the data, ensuring comparable distributions. 
In this work, we measure angular correlations. For this reason, we are interested in producing simulations that mimic the data's angular distribution as accurately as possible. We achieve this by generating a distribution of events using the inverse transform method, which produces pseudo-random numbers based on a probability density function (pdf) \cite{devroye:1986}. Since the simulations need to describe our background scenario, which consists of the absence of any UHECRs  signal from known sources, we define the pdf used to sample our pseudo-data as the convolution of two pdfs: one describing the PAO angular acceptance (displayed in
Fig.~2 of the main paper) and the other accounting for the measured dipole anisotropy at energies above 32\,EeV \citep{pao_anisotropy_2024}. Despite the measurement of a dipole anisotropy with significance above 5$\sigma$ at energies above 8~EeV, and only some mild evidence of anisotropies at higher energies, we model the background according to the most recently measured anisotropy above 32~EeV \cite{pao_anisotropy_2024}. This choice is motivated by the energy range of the data sample utilized in the analysis, which comprises only events with energies above 32~EeV. In \autoref{fig:simulation}, we present a comparison between the simulated distributions and the data. The two distributions are in good agreement, with their ratios averaging at $\sim1$ over the whole analyzed coordinate range. We note a slight undercoverage of the data between a declination from $\sim-50^\circ$ to $\sim-40^\circ$. The disagreement might be related to the established excess observed in the Centaurus region, both by PAO \cite{PAO_Anisotropy2022_ApJ} and in this work, and is located in that exact declination range. 

The significance of the results is evaluated by running a correlation analysis on more than 30000 pseudo-experiments and comparing the correlation results from the data with the distribution of correlation outcomes from the simulations. The p$_\text{value}$ is defined as the percentage of the distribution that provides a result at least as extreme as the one observed in the data. 

The simulations were also used to calculate the uncertainty bands for the local excesses of each source by counting the number of events within a search radius $\psi$ at each analyzed source location across a few hundred pseudo-experiments. 

\newpage
\section{List of Analyzed Sources}
In \autoref{sources} we list all sources reported in \cite{Kormendy_2013, Saglia_2016} that were considered possible UHECR candidates in this work, also motivated by other works (e.g., 
\citep{Torres2002}).
\begin{longtable*}{lcccccc}
\caption{\textbf{List of galaxies with reliable SMBH measurements.} Sources
are ordered by ascending declination. For each source, we list the name, redshift, equatorial coordinates (J2000 equinox), distance in Mpc, and the SMBH's mass in solar masses. The Active galaxies are marked with a $\bullet$ in the last column. The sources excluded from the analysis due to the PAO field of view are marked with a *. We highlight the Galactic Center, SgrA*, which is not included in the analysis reported in the main test but presented in some additional results in the supplementary material.}\label{sources}\\
\toprule
Object & Redshift & R.A. (deg) & Dec. (deg) & D (Mpc) & M$_{\text{SMBH}}$ (M$_\odot$) & Activity \\
\midrule
\endfirsthead
\toprule
Object & Redshift & R.A. (deg) & Dec. (deg) & D (Mpc) & M$_{\text{SMBH}}$ (M$_\odot$) & Activity \\
\midrule
\endhead
\midrule
\multicolumn{7}{r}{Continued on next page} \\
\midrule
\endfoot
\bottomrule
\endlastfoot
Circinus Galaxy & 0.001 & 213.29 & $-$65.34 & 2.8 & 1.14e+06 & $\bullet$ \\
NGC\,4945      & 0.002 & 196.36 & $-$49.47 & 3.6 & 1.35e+06 & $\bullet$ \\
NGC\,6861      & 0.009 & 301.83 & $-$48.37  & 28.7 & 2.10e+09 & $\bullet$ \\
NGC\,5516      & 0.014 & 213.98 & $-$48.11  & 55.3 & 3.69e+09 &   \\
Cen\,A      & 0.002 & 201.37 & $-$43.02  & 3.6 & 5.69e+07 & $\bullet$ \\
NGC\,4751      & 0.007 & 193.21 & $-$42.66  & 32.8 & 1.71e+09 &   \\
NGC\,7582      & 0.005 & 349.60 & $-$42.37  & 22.3 & 5.51e+07 & $\bullet$ \\
NGC\,1316      & 0.006 & 50.67 & $-$37.21  & 20.9 & 1.69e+08 & $\bullet$ \\
IC\,1459       & 0.006 & 344.29 & $-$36.46  & 28.9 & 2.48e+09 & $\bullet$ \\
NGC\,1399      & 0.005 & 54.62 & $-$35.45  & 20.9 & 8.81e+08 & $\bullet$ \\
NGC\,1374      & 0.005 & 53.82 & $-$35.23  & 19.6 & 5.90e+08 &   \\
NGC\,5419      & 0.014 & 210.91 & $-$33.98  & 56.2 & 7.24e+09 & $\bullet$ \\
IC\,4296 & 0.012 & 204.16 & $-$33.97  & 49.2 & 3.74e+09 & $\bullet$ \\
IC\,2560       & 0.010 & 154.08 & $-$33.56  & 37.2 & 5.01e+06 & $\bullet$ \\
\textbf{SgrA*} & 0.000 & 266.42 & $-$29.00  & 0.008 & 4.30e+06 &  \\
NGC\,3923      & 0.006 & 177.76 & $-$28.81  & 20.8 & 2.81e+09 &   \\
NGC\,5328      & 0.016 & 208.22 & $-$28.49  & 64.4 & 4.75e+09 &   \\
NGC\,3585      & 0.005 & 168.32 & $-$26.75  & 20.5 & 3.29e+08 &   \\
NGC\,1398      & 0.005 & 54.72 & $-$26.34  & 24.8 & 1.08e+08 &   \\
NGC\,3393      & 0.013 & 162.10 & $-$25.16  & 49.2 & 1.57e+07 & $\bullet$ \\
NGC\,1332      & 0.005 & 51.57 & $-$21.34  & 22.7 & 1.47e+09 &   \\
NGC\,3091      & 0.013 & 150.06 & $-$19.64  & 53.0 & 3.72e+09 &   \\
NGC\,5018      & 0.009 & 198.25 & $-$19.52  & 40.6 & 1.05e+08 &   \\
NGC\,1300      & 0.005 & 49.92 & $-$19.41  & 21.5 & 7.55e+07 &   \\
NGC\,1407      & 0.006 & 55.05 & $-$18.58  & 29.0 & 4.65e+09 &   \\
NGC\,5077      & 0.009 & 199.88 & $-$12.66  & 38.7 & 8.55e+08 & $\bullet$ \\
MESSIER\,104   & 0.004 & 190.00 & $-$11.62  & 9.9 & 6.65e+08 & $\bullet$ \\
ABELL\,1836    & 0.038 & 210.42 & $-$11.61  & 152.4 & 2.66e+09 &   \\
NGC\,4699      & 0.005 & 192.26 & $-$8.66  & 18.9 & 1.76e+08 &   \\
NGC\,3115      & 0.002 & 151.31 & $-$7.72  & 9.5 & 8.97e+08 &   \\
NGC\,4697      & 0.004 & 192.15 & $-$5.8 0 & 12.5 & 2.02e+08 &   \\
NGC\,2974      & 0.006 & 145.64 & $-$3.7 0 & 21.5 & 1.70e+08 & $\bullet$ \\
NGC\,0307      & 0.013 & 14.14 & $-$1.77  & 52.8 & 4.00e+08 &   \\
NGC\,1194      & 0.014 & 45.95 & $-$1.1 0 & 58.0 & 7.08e+07 & $\bullet$ \\
NGC\,1068   & 0.004 & 40.67 & $-$0.01  & 15.9 & 8.39e+06 & $\bullet$ \\
NGC\,5846      & 0.006 & 226.62 & 1.61  & 24.9 & 1.10e+09 & $\bullet$ \\
NGC\,5845      & 0.006 & 226.50 & 1.63  & 25.9 & 4.87e+08 &   \\
NGC\,5813      & 0.007 & 225.30 & 1.70 & 32.2 & 7.10e+08 & $\bullet$ \\
NGC\,1550      & 0.012 & 64.91 & 2.41  & 52.5 & 3.87e+09 &   \\
NGC\,5576      & 0.005 & 215.27 & 3.27  & 25.7 & 2.73e+08 & $\bullet$ \\
NGC\,2960      & 0.017 & 145.15 & 3.58  & 67.1 & 1.08e+07 & $\bullet$ \\
NGC\,4261      & 0.007 & 184.85 & 5.82  & 32.4 & 5.29e+08 & $\bullet$ \\
IC\,1481       & 0.020 & 349.85 & 5.91  & 89.9 & 1.49e+07 & $\bullet$ \\
NGC\,4342      & 0.003 & 185.91 & 7.05  & 22.9 & 4.53e+08 &   \\
NGC\,4526      & 0.002 & 188.51 & 7.7 0 & 16.4 & 4.51e+08 &   \\
MESSIER\,049   & 0.003 & 187.44 & 8.00 & 16.7 & 2.54e+09 & $\bullet$ \\
NGC\,7619      & 0.013 & 350.06 & 8.21  & 53.9 & 2.30e+09 &   \\
NGC\,0524      & 0.008 & 21.20 & 9.54  & 24.2 & 8.67e+08 &   \\
NGC\,4596      & 0.006 & 189.98 & 10.18  & 16.5 & 7.67e+07 &   \\
NGC\,0821      & 0.006 & 32.09 & 10.99  & 23.4 & 1.65e+08 &   \\
NGC\,4564      & 0.004 & 189.11 & 11.44  & 15.9 & 8.81e+07 &   \\
MESSIER\,060   & 0.004 & 190.92 & 11.55  & 16.5 & 4.72e+09 &   \\
NGC\,4371      & 0.003 & 186.23 & 11.70 & 16.9 & 7.00e+06 &   \\
MESSIER\,096   & 0.003 & 161.69 & 11.82  & 10.6 & 7.66e+06 &   \\
NGC\,4486A & 0.003 & 187.67 & 12.27  & 18.4 & 1.44e+07 &   \\
MESSIER\,087   & 0.004 & 187.71 & 12.39  & 16.7 & 6.15e+09 & $\bullet$ \\
NGC\,4486B & 0.005 & 187.63 & 12.49  & 16.3 & 6.00e+08 &   \\
MESSIER\,089   & 0.001 & 188.92 & 12.56  & 15.3 & 5.00e+08 & $\bullet$ \\
MESSIER\,105   & 0.003 & 161.96 & 12.58  & 10.7 & 4.16e+08 &   \\
NGC\,3384      & 0.002 & 162.07 & 12.63  & 11.5 & 1.08e+07 &   \\
NGC\,4388      & 0.008 & 186.44 & 12.66  & 16.5 & 7.31e+06 & $\bullet$ \\
MESSIER\,084   & 0.003 & 186.27 & 12.89  & 18.5 & 9.25e+08 & $\bullet$ \\
MESSIER\,066   & 0.002 & 170.06 & 12.99  & 10.5 & 8.49e+07 & $\bullet$ \\
NGC\,4473      & 0.007 & 187.45 & 13.43  & 15.2 & 9.00e+07 &   \\
NGC\,3489      & 0.002 & 165.08 & 13.9 0 & 12.0 & 5.94e+06 &   \\
NGC\,4459      & 0.004 & 187.25 & 13.98  & 16.0 & 6.96e+07 &   \\
NGC\,3377      & 0.002 & 161.93 & 13.99  & 11.0 & 1.78e+08 &   \\
MESSIER\,088   & 0.008 & 188.00 & 14.42  & 16.5 & 2.00e+07 & $\bullet$ \\
NGC\,3607      & 0.003 & 169.23 & 18.05  & 22.6 & 1.37e+08 & $\bullet$ \\
NGC\,3608      & 0.004 & 169.25 & 18.15  & 22.8 & 4.65e+08 &   \\
MESSIER\,085   & 0.002 & 186.35 & 18.19  & 17.9 & 1.30e+07 &   \\
NGC\,3227      & 0.004 & 155.88 & 19.87  & 23.8 & 2.10e+07 & $\bullet$ \\
NGC\,3842      & 0.021 & 176.01 & 19.95  & 92.2 & 9.09e+09 &   \\
MESSIER\,064   & 0.001 & 194.18 & 21.68  & 7.3 & 1.56e+06 & $\bullet$ \\
NGC\,7052      & 0.016 & 319.64 & 26.45  & 70.4 & 3.96e+08 & $\bullet$ \\
NGC\,7768      & 0.027 & 357.74 & 27.15  & 116.0 & 1.34e+09 &   \\
NGC\,6264      & 0.034 & 254.32 & 27.85  & 147.6 & 3.08e+07 & $\bullet$ \\
NGC\,3414      & 0.005 & 162.82 & 27.98  & 25.2 & 2.51e+08 &   \\
NGC\,4886 & 0.022 & 195.03 & 27.98  & 102.0 & 2.08e+10 &   \\
NGC\,3245      & 0.004 & 156.83 & 28.51  & 21.4 & 2.39e+08 & $\bullet$ \\
NGC\,6086      & 0.031 & 243.15 & 29.48  & 138.0 & 3.74e+09 & $\bullet$ \\
NGC\,7457      & 0.003 & 345.25 & 30.14  & 12.5 & 9.00e+06 &   \\
NGC\,2778      & 0.007 & 138.10 & 35.03  & 23.4 & 1.45e+07 &   \\
NGC\,1023      & 0.002 & 40.10 & 39.06  & 10.8 & 4.13e+07 &   \\
NGC\,4151      & 0.003 & 182.64 & 39.41  & 20.0 & 6.50e+07 & $\bullet$ \\
Cygnus\,A      & 0.056 & 299.87 & 40.73  & 242.7 & 1.30e+09 & $\bullet$ \\
MESSIER\,032   & $-$0.001 & 10.67 & 40.87  & 0.8 & 2.45e+06 &   \\
MESSIER\,094   & 0.001 & 192.72 & 41.12  & 5.0 & 6.77e+06 & $\bullet$ \\
MESSIER\,031   & $-$0.001 & 10.68 & 41.27  & 0.8 & 1.43e+08 &   \\
NGC\,1277      & 0.017 & 49.96 & 41.57  & 73.0 & 1.70e+10 &   \\
NGC\,6323      & 0.026 & 258.33 & 43.78  & 113.4 & 1.01e+07 & $\bullet$ \\
MESSIER\,106*   & 0.002 & 184.74 & 47.3 0 & 7.3 & 3.78e+07 & $\bullet$ \\
NGC\,4026*      & 0.003 & 179.85 & 50.96  & 13.3 & 1.80e+08 &   \\
NGC\,3998*      & 0.003 & 179.48 & 55.45  & 14.3 & 8.45e+08 & $\bullet$ \\
NGC\,3079*      & 0.004 & 150.49 & 55.68  & 15.9 & 2.50e+06 & $\bullet$ \\
NGC\,2549*     & 0.004 & 124.74 & 57.8 0 & 12.7 & 1.45e+07 &   \\
UGC\,03789*     & 0.011 & 109.88 & 59.36  & 49.9 & 9.65e+06 &   \\
NGC\,3945*      & 0.004 & 178.31 & 60.68  & 19.5 & 8.80e+06 &   \\
NGC\,2273*      & 0.006 & 102.54 & 60.85  & 29.5 & 8.61e+06 & $\bullet$ \\
MESSIER\,081*   & 0.000 & 148.89 & 69.07  & 3.6 & 6.50e+07 & $\bullet$ \\
NGC\,2787*      & 0.002 & 139.83 & 69.2 0 & 7.5 & 4.07e+07 & $\bullet$ \\
NGC\,4291*      & 0.006 & 185.08 & 75.37  & 26.6 & 9.78e+08 & $\bullet$ \\
NGC\,2748*      & 0.005 & 138.43 & 76.48  & 23.4 & 4.44e+07 &   \\
NGC\,6251*      & 0.025 & 248.13 & 82.54  & 108.4 & 6.14e+08 & $\bullet$ \\
\bottomrule
\end{longtable*}

\section{Accounting for the look-elsewhere effect}
\begin{figure}
    \centering
    \includegraphics[width=0.5\linewidth]{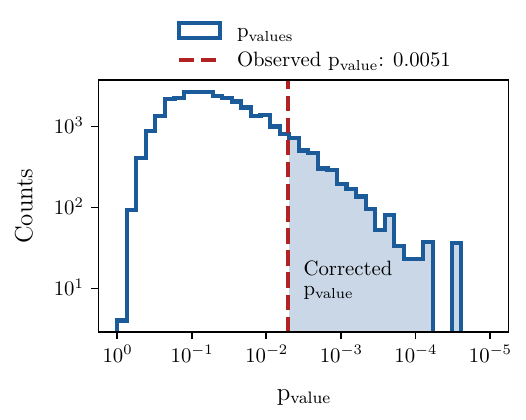}
    \caption{Distribution of the minimum p-values across all tested cases for all the performed pseudo-experiments. The dashed red line represents the observed minimum p$_\text{value}$ in the data, and the shaded area represents the p$_\text{value}$ corrected for the look-elsewhere effect.}
    \label{fig:trial_correction}
\end{figure}

When testing several hypotheses, the significance of the result needs to be corrected for the look-elsewhere effect. In our case, we search for correlations between UHECRs and the sources in our list over 30 search radii, $\psi$, and 3 source classes (AGN, quiescent galaxies, and all sources). Since the hypotheses are highly correlated, the correction is performed based on the simulations. The final significance is estimated as follows: we run the analysis on each pseudo-dataset and construct the distribution of the lowest p$_\text{value}$ across all tested cases. The final significance corresponds to the fraction of the distribution with values smaller than the observed result. This is shown in \autoref{fig:trial_correction}, where the histogram represents the distribution of the lowest p$_\text{value}$ obtained across all tested cases for each pseudo-experiment produced for this analysis, the dashed line is the observed minimum p$_\text{value}$, and the colored area, corresponding to $\sim\!10\%$ of the total distribution, is the p$_\text{value}$ corrected for the look-elsewhere effect. 

\section{More Results}

\subsection{Energy and confusion checks}
Looking at the energies of the events contributing to the excess, no correlation between event energy and their proximity to the source is observed.
Additionally, to assess potential class confusion, we inject signal-like events distributed as 2D Gaussians centered on source coordinates and repeat the analysis for all pairwise combinations of source class and test (Tab.~\ref{tab:results}).
The observed fractions of events contributing to the excesses are reproduced by 
$\sim\!300$ injected events on AGN or $\sim\!200$--$500$ on quiescent nuclei, with mild dependence on search radius, and $\sim\!300$ events in either class reproduce the combined excess.
Within the current exposure, the modest excess cannot be uniquely attributed to a single source class. In summary, local excesses are observed in both source classes, but are consistent with isotropy after correcting for the look-elsewhere effect. Magnetic deflections, limited statistics, and the spatial overlap between AGN and quiescent nuclei in the Centaurus region prevent an unambiguous attribution; current data are consistent with a mixed or class-independent contribution, as expected if UHECR acceleration is governed by the gravity-limited mechanism rather than AGN activity per se.

\begin{table}[ht]
  \centering
  \caption{Signal-injection results. 
  Median number of injected events required to 
  reproduce the observed excess fractions for all pairwise 
  combinations of injected class (columns) and tested correlation 
  (rows).}
  \label{tab:results}
  \begin{tabular}{l|ccc}
    \toprule
    & \multicolumn{3}{c}{$N_{\rm inj}$ signal events} \\
    \midrule
    Correlation & AGN & Quiescent & All \\
    \midrule
    AGN ($\psi\!=\!22^\circ$)      & 200 & 500 & 300 \\
    Quiescent ($\psi\!=\!20^\circ$) 
                                    & 300 & 200 & 200 \\
    All ($\psi\!=\!24^\circ$)       & 300 & 300 & 300 \\
    \bottomrule
  \end{tabular}
\end{table}

\subsection{Cross-checks of the results including Sgr~A*}

Our own galaxy also hosts a SMBH, Sgr~A*, having a mass in the allowed range of masses for the analysis presented in this work. As explained in the main text, it was excluded from the analysis due to its proximity to Earth and the lack of evidence for any UHECRs originating from its direction. However, given that SgrA* meets the requirements for hosting gravity-limited acceleration, we also present the results of the analysis when it is included in the list as a quiescent galaxy.

\begin{figure}
    \centering
    \includegraphics[width=.8\linewidth]{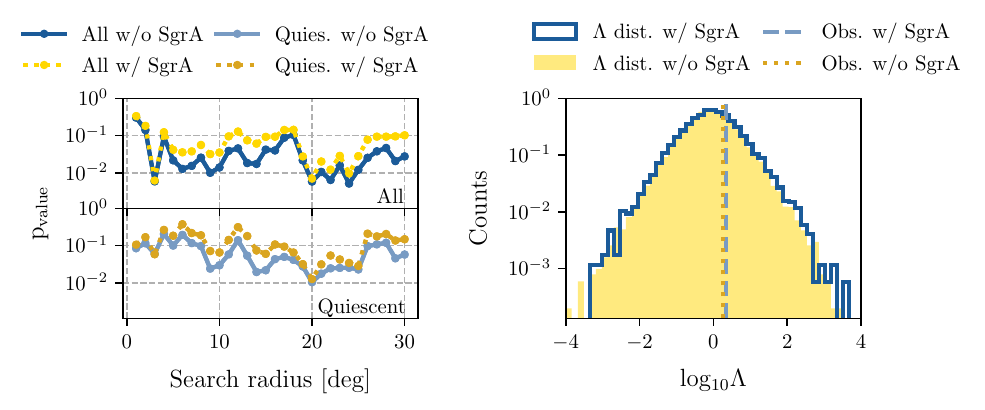}
    \caption{Left: Local p$_{\rm values}$ of the analysis for the quiescent galaxies (Quies., bottom left panel) and all the sources (All, upper left panel) as a function of the tested search radii, both including Sgr~A* to the analysis (yellows) and not (blues). The AGN class is not shown as it's not affected by the addition of a quiescent candidate. Right: $\log_{10}\Lambda$ distributions, both including SgrA* to the analysis (yellows) and not (blues), and the corresponding observed values, which fall close to the center of the distribution, indicating no preference for either of the classes. }
    \label{fig:pvals_with_SgrA}
\end{figure}

In \autoref{tab:results_withSgrA*} and \autoref{fig:pvals_with_SgrA} we display the correlation results obtained when including Sgr~A* in the list and compare them with those obtained without it. The correlation results for the AGN class are unaffected, as Sgr~A* is a quiescent galaxy; the results of the other two tests change due to the updated source list. 
The addition of Sgr~A* slightly lowers the significance of the excess for the quiescent galaxies, but the excess stays at a search radius $\psi = 20^\circ$. In the case of the complete list of sources, instead, the most significant excess is found at an angular distance of $\psi = 3^\circ$ with a significance lower but not too different from the one found for a search radius of $24^\circ$ (the most significant excess when excluding Sgr~A* from the analysis). 

Here, we also present the results of a likelihood-ratio test to determine whether there is a class preference. 
The test statistic is defined as
\begin{equation}
\Lambda = \frac{\min_\psi p_{\rm AGN}}{\min_\psi 
p_{\rm quiescent}}.
\end{equation}
The data yield $\log_{10}\Lambda\!=\!0.25$ ($\Lambda\!=\!1.81$) when Sgr~A* is excluded, and $\log_{10}\Lambda = 0.33$ ($\Lambda = 2.18$) when it's included.
Both cases are fully consistent with the median of the background 
distribution and statistically not significant
(Fig.~\ref{fig:pvals_with_SgrA}, right): no preference 
for either source class is established with current data.
\begin{table}[ht]
  \centering
  \caption{Comparison of the results including Sgr~A* to the correlation analysis (left columns, With Sgr~A*) and the ones presented in the main paper (right columns, 
  without Sgr~A*). The significances reported here are all \textit{local}, i.e., do not account for the look-elsewhere effect.}\label{tab:results_withSgrA*}
    \begin{tabular}{c|cc|cc}
    \toprule
    \multirow{2}{*}{Source Class} &\multicolumn{2}{c|}{With Sgr~A*} & \multicolumn{2}{c}{Without Sgr~A*}\\
    & $\psi$ [deg] & p$_{\text{value}}$ & $\psi$ [deg] & p$_{\text{value}}$ \\
    \midrule
    AGN & 22$^\circ$ & 0.0058 (2.76$\sigma$) & 22$^\circ$ & 0.0058 (2.76$\sigma$) \\
    Quiescent Galaxies & 20$^\circ$ & 0.012 (2.50$\sigma$)& 20$^\circ$ & 0.010 (2.56$\sigma$) \\
    All Sources & 3$^\circ$ & 0.006 (2.73$\sigma$) & 24$^\circ$ & 0.0051 (2.79$\sigma$)\\
    \bottomrule
    \end{tabular}%
\end{table}

Looking at the distribution of the p-values in \autoref{fig:pvals_with_SgrA}, it is not surprising that one single source is able to move the excess from $24^\circ$ to $3^\circ$. In fact, in both cases, we notice the presence of two distinguished excesses in the p-value distribution peaking at very similar p-values. Additionally, we note that the addition of the Galactic Center,  given that most of the sources are distributed among the Supergalactic Plane (see Figure 2 of the main paper), adds information about a portion of the Sky which is otherwise not considered in the analysis, slightly reducing the correlation between the analyses and increasing the probability of getting different outcomes. 

We also show the local excesses for the sources contributing to the most significant excess. We notice that the most significant local excesses are concentrated around the Centaurus region, once again showing 3 sources with excesses above $2\sigma$ relative to the median expectation. These three AGN are the same mentioned in the main paper: NGC\,4945 ($\delta=-49.5^\circ$), Cen\,A ($-43.0^\circ$), and NGC\,5419 ($-34.0^\circ$). 
\begin{figure}
    \centering
    \includegraphics[width=0.45\linewidth]{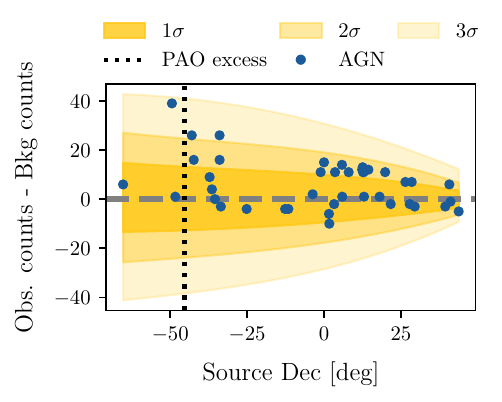}
    \caption{Per-source excess relative to the median background. The source's excesses are represented by the blue dots, the shaded bands correspond to the 1$\sigma$, 2$\sigma$, 3$\sigma$ background contours. Three AGN (NGC\,4945, Cen\,A, NGC\,5419) exceed $2\sigma$. The dashed line represents the declination of the most significant excess found by PAO, which is $-45.4^\circ$~\cite{PAO_SGP_2025}.}
    \label{fig:dev_from_background}
\end{figure}

\subsection{The second most significant excess}

\begin{figure}
    \centering
    \includegraphics[width=0.45\linewidth]{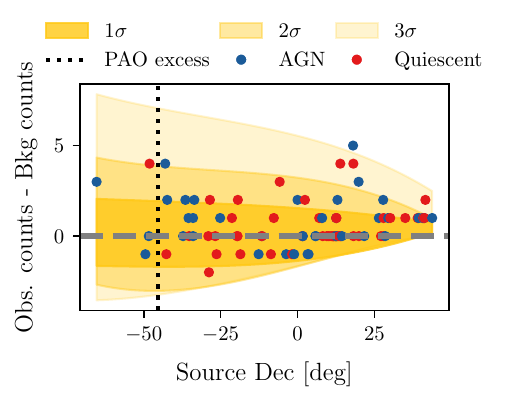}
    \caption{Per-source excess relative to the median background. The source's excesses are represented by the blue dots, the shaded bands correspond to the 1$\sigma$, 2$\sigma$, 3$\sigma$ background contours. Five sources exceed $2\sigma$. The dashed line represents the declination of the most significant excess found by PAO, which is $-45.4^\circ$~\cite{PAO_SGP_2025}.}
    \label{fig:dev_from_background_3deg}
\end{figure}

Finally, it is interesting to note that in all the tested cases, we observe a prominent excess at a search radius of $3^\circ$, which also becomes the most significant when testing all sources, including Sgr~A*. In \autoref{fig:dev_from_background_3deg}, we show the number of events that exceed the background expectation in this specific case. We notice that two sources, the quiescent galaxy NGC~5516 ($\delta = -48.11^\circ$) and the AGN NGC~5128 ($\delta = -43.02^\circ$), which present mild excesses, are located in the Centaurus region, which appears to have a number of associated events above the median expected background also for small search radii. Additionally, three other sources, the quiescent galaxies NGC~3489 ($\delta = 13.90^\circ$)  and NGC~3608 ($\delta = 18.15^\circ$), and the active galaxy NGC~3607 ($\delta = 18.05^\circ$) also emerge from the background more significantly. 

However, given that the analysis aimed to identify the most significant excess among the tested cases, we refrain from drawing any conclusions regarding this excess. 

\end{document}